\documentclass{Interspeech2024}
\usepackage{adjustbox}
\usepackage{float}

\interspeechcameraready 

\title{TalTech-IRIT-LIS Speaker and Language Diarization Systems\\for 
DISPLACE 2024}

\name[affiliation={1}]{Joonas}{Kalda}
\name[affiliation={1}]{Tanel}{Alumäe}
\name[affiliation={2}]{Martin}{Lebourdais}
\name[affiliation={2}]{\\Hervé}{Bredin}
\name[affiliation={3}]{Séverin}{Baroudi}
\name[affiliation={3}]{Ricard}{Marxer}

\address{
  $^1$Tallinn University of Technology, Estonia \hspace{0.1cm}
    $^2$IRIT, Universit\'{e} de Toulouse, CNRS, Toulouse, France \\
    $^3$Universit\'{e} de Toulon, Aix Marseille Univ, CNRS, LIS, Toulon, France}
\email{firstname.lastname@\{taltech.ee, irit.fr, lis-lab.fr\}}


\usepackage{pifont}

\newcommand{\cmark}{\ding{51}}%
\newcommand{\xmark}{\ding{55}}%

\begin{document}

\maketitle

\begin{abstract}
This paper describes the submissions of team TalTech-IRIT-LIS to the DISPLACE 2024 challenge. Our team participated in the speaker diarization and language diarization tracks of the challenge. In the speaker diarization track, our best submission was an ensemble of systems based on the \textit{pyannote.audio} speaker diarization pipeline utilizing powerset training and our recently proposed PixIT method that performs joint diarization and speech separation. We improve upon PixIT by using the separation outputs for speaker embedding extraction. Our ensemble achieved a diarization error rate of 27.1\% on the evaluation dataset. In the language diarization track, we fine-tuned a pre-trained Wav2Vec2-BERT language embedding model on in-domain data, and clustered short segments using AHC and VBx, based on similarity scores from LDA/PLDA. This led to a language diarization error rate of 27.6\% on the evaluation data. Both results were ranked first in their respective challenge tracks.
\end{abstract}

\section{Introduction}

Speaker diarization is the task of dividing an audio recording into segments based on the speaker identity. The conventional method for tackling this is a multi-stage approach that joins speaker segmentation, local speaker embeddings, and clustering \cite{landini2022bayesian}. This approach struggles with overlap-heavy speech, a domain that is better suited for end-to-end neural diarization (EEND) \cite{fujitaEndtoEndNeuralSpeaker2019, fujitaEndtoEndNeuralSpeaker2019a}. On the other hand, EEND is data-hungry and has the issue of mispredicting the number of speakers. This has motivated a hybrid approach that replaces the speaker segmentation step of the multi-stage approach with local EEND \cite{kinoshitaIntegratingEndtoendNeural2021b}.

Language diarization is the less-studied task of segmenting a recording by the spoken language. It is used as the first step in processing multilingual code-switched speech. Inspired by speaker diarization, both multi-stage \cite{baghel2024summary} and end-to-end neural \cite{mishra2021spoken} approaches have been used to solve this task.

\begin{figure}[tb]
    \centering
    \includegraphics[width=1\linewidth]{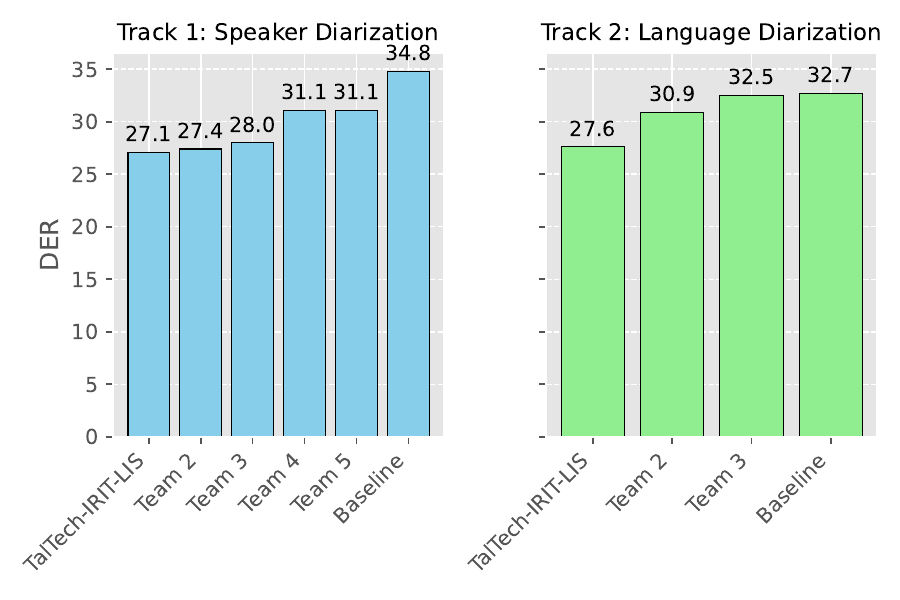}
    \caption{The results of top-performing teams on DISPLACE 2024 evaluation data.}
    \label{fig:rankings}
\end{figure}

The DISPLACE 2024 Challenge is centered on advancing research in the domains of speaker and language diarization, as well as automatic speech recognition (ASR), within multilingual and multi-accent environments \cite{kalluri24_interspeech}. The challenge emphasizes the utilization of realistic speech data, characteristically featuring frequent language switches by speakers at both sentence and phrase levels. DISPLACE 2024 is structured around three evaluation tracks: speaker diarization, language diarization, and ASR.

The dataset for the first two tracks comprises far-field, multi-party multilingual conversational speech recordings, featuring speakers who engage in code-mixing or code-switching across multiple languages. The development set for these tracks consists of 35 recordings, summing up to nearly 20 hours of audio. The evaluation set encompasses 32 recordings, totaling almost 18 hours. Each recorded conversation, lasting around 30 to 60 minutes, involves 3-5 participants fluent in various Indian languages as well as English (with an Indian accent).

For the third track, dedicated to speech recognition, a separate development dataset is provided. This dataset includes 8 recordings, each segmented into single-language regions. The segments are labeled with the corresponding language and accompanied by an orthographic transcript.

Participants in the challenge are permitted to employ any publicly available or proprietary datasets for training and refining their diarization systems. This includes leveraging development data from other tracks within the challenge. These development sets can be utilized for model training and hyper-parameter optimization. The performance of systems in Tracks 1 and 2 is evaluated based on the diarization error rate (DER), with overlap and without forgiveness collar.

Our team participated in Tracks 1 and 2 of the challenge. Figure \ref{fig:rankings} shows that we outperformed other teams in both tracks.

\section{Track 1: Speaker diarization}

\subsection{Methods}

\subsubsection{Powerset training}
\label{section:poweset}

Our first standalone system is based on the same approach as the submission \#5 of the \textit{pyannote} team at VoxSRC 2023~\cite{voxsrc2023}. This hybrid approach consists of local end-to-end neural speaker segmentation
on a few-second sliding window, neural speaker embedding of
each speaker of each window, and agglomerative hierarchical clustering (AHC). The backbone of our local speaker segmentation model is a WavLM-base model \cite{chen_wavlm_2022} pre-trained from scratch on a compound dataset consisting of AISHELL \cite{aishell_2017}, AliMeeting \cite{yu_m2met_2022}, AMI \cite{carletta2005ami}, AVA-AVD \cite{xu2022ava}, DIHARD \cite{ryant2021dihard}, Ego4D \cite{grauman2022ego4d}, MSDWild \cite{liu2022msdwild}, REPERE \cite{giraudel_repere_2012}, and VoxConverse 0.3.0 \cite{Chung_2020_voxconverse} which is applied on a 10-second sliding window with a stride of 1 second. The 8th layer of this WavLM is fed into an LSTM-based network. The WavLM and the LSTM-based network consist of 94.4M parameters and 2.1M parameters respectively. The training uses powerset multi-label cross-entropy loss \cite{plaquetPowersetMulticlassCross2023} with $K_{\max}=3$ speakers. Speaker embeddings were extracted using the pre-trained ResNet34 model from the WeSpeaker toolkit  \cite{wang2022wespeaker}.


\subsubsection{PixIT}

We also experimented with our recently proposed PixIT method, combining permutation invariant training (PIT) for speaker diarization and mixture invariant training (MixIT) for speech separation \cite{kalda2024pixit}. For PixIT, the multitask loss is defined as $\mathcal{L}_{\mathrm{PixIT}} = \lambda \mathcal{L}_{\mathrm{PIT}} + 
(1-\lambda) \mathcal{L}_{\mathrm{MixIT}}$. It is calculated on pairs of mixtures extracted from the same recording environments so that they contain disjoint sets of speakers and the combined number of speakers is at most $K_{\max}$. The mixtures are added together to create mixtures of mixtures (MoMs). $\mathcal{L}_{\mathrm{MixIT}}$ is calculated only on the MoMs while $\mathcal{L}_{\mathrm{PIT}}$ also utilizes the original mixtures. The local joint model is based on the TasNet architecture \cite{luoTasNetTimedomainAudio2018}. The feature encoder concatenates the outputs of the pre-trained WavLM-large \cite{chen_wavlm_2022} model and a 1-D convolutional encoder. The masking network outputs $K_{\max}$ masks which are then independently processed by either a 1-D convolutional decoder or a fully connected neural network for local speech separation or speaker diarization respectively. 

To perform global speaker diarization, the speaker diarization branch of the local joint model is used in the same pipeline as in Section \ref{section:poweset}. The only difference is that for speaker embeddings we used a pre-trained ECAPA-TDNN model \cite{desplanques2020ecapa} available in \cite{ravanelli2021speechbrain}.

We experimented with multiple improvements to the original PixIT system. First, we utilized separated sources output by the joint model for speaker embedding extraction instead of the original audio. This allows for additional information from the overlapped regions and further integrates the two tasks. A potential downside is that separation outputs can include artifacts the speaker embedding model has not seen during its training. Second, we used a DPTNet \cite{chen20_dptnet} instead of a DPRNN \cite{luoDualpathRNNEfficient2020} as the masking network which was shown to perform better at speech separation albeit on synthetic data. We kept the hyperparameters the same as in the original work. Finally, to improve the quality of the local speaker embeddings we increased the length of the sliding window from 5 to 10 seconds while increasing the stride of the convolutional encoder two-fold.

The total number of parameters for the PixIT model is 319M when using a DPRNN and 324M when using a DPTNet.

\subsection{Results}

For fine-tuning our speaker diarization systems, we divided the DISPLACE 2024 development set further into train and development splits with the latter containing the recordings M030, B022, M019, and B034. Accordingly, we will only report results on the evaluation dataset of the challenge.

The performance on the evaluation dataset for our PixIT-based systems is detailed in Table \ref{tab:pixit}. Optimizing for the DER on the development data, we found $\lambda = 0.1$ to perform the best. Using the separated sources predicted by the joint model for extracting speaker embeddings instead of the original audio yields an improvement in DER from 30.1\% to 29.4\%. This shows that the additional information extracted from the overlapped regions outweighs the negative effect of the presence of artifacts in the separated sources. An additional 7.8\% relative improvement is achieved by replacing the DPRNN with a DPTNet as the masking network. The superior performance of DPTNet thus extends to the case of shared training on real data. Lastly, extending the sliding window length to 10 seconds further improves DER by a relative 1.7\%.

\begin{table}[tb]
    \centering
    \caption{DERs (\%) obtained on Track 1 evaluation data for different configurations of the PixIT method. $\ast$ denotes submissions made during the post-evaluation phase of the competition.}
    \setlength{\tabcolsep}{3pt}
    \adjustbox{max width=\linewidth}{
    \begin{tabular}{ll}    
    \hline
    Submission & Eval \\\hline
    \qquad DISPLACE 2024 baseline & 34.76 \\
    \#1\quad Original PixIT system with $\lambda=0.1$ & 30.05 \\
    \#2\quad \#1 + embeddings from separated sources & 29.44 \\
    \#3\quad \#2 + DPTNet as the masking network & $27.15^\ast$\\
    \#4\quad \#3 + 10s sliding window  & $26.70^\ast$\\\hline
    \end{tabular}}
    
    \label{tab:pixit}
\end{table}

The results of our systems using powerset training and ensemble methods are shown in Table \ref{tab:doverlap}. Fine-tuning the powerset system allowed us to get from 30.6\% down to 27.3\%  DER on the evaluation data. We also experimented with constraining the maximal number of speakers in clustering to either 5, 6, or 7. The last case yields slight improvements to DER while others perform worse than the unconstrained system. Finally, we use greedy DOVER-Lap~\cite{rajDOVERLapMethodCombining2020} to combine the PixIT system with various powerset systems. We found the best results from choosing the unconstrained fine-tuned version and the fine-tuned version constrained to a maximum of 5 speakers. This is likely because the variation in outputs is the greatest for that pair of systems.

\begin{table}[tb]
    \centering
    \caption{DERs (\%) obtained on Track 1 evaluation dataset for different system configurations. Our best-performing system for Phase 1 of the competition is in bold.}
    \adjustbox{max width=\linewidth}{
    \begin{tabular}{lc}
    \hline
       Submission   & Eval\\\hline
        \qquad DISPLACE 2024 baseline & 34.76 \\
        \#2\quad PixIT              & 29.44\\
        \#5\quad powerset off-the-shelf             & 30.57\\
        \#6\quad powerset fine-tuned              & 27.34\\
        \#7\quad powerset fine-tuned, $\textit{max\_speakers}=5$            & 29.09\\
        \#8\quad powerset fine-tuned, $\textit{max\_speakers}=6$            & 28.35\\
        \#9\quad powerset fine-tuned, $\textit{max\_speakers}=7$           & 27.29\\
        \#10\hspace{0.25em} DOVER-Lap of \#2, \#6 and \#9 & 27.27\\
        \#11\hspace{0.25em} DOVER-Lap of \#2, \#6, \#7, \#8 and \#9 & 27.12\\
        \#12\hspace{0.25em} DOVER-Lap of \#2, \#6 and \#7 & \textbf{27.08}\\\hline
        
    \end{tabular}}
    
    \label{tab:doverlap}
\end{table}

\subsection{Runtime performance}

The powerset system was fine-tuned using a single V100 GPU for approximately 1h. On the same hardware, it takes 10m30s to process the DISPLACE 2024 evaluation set. PixIT systems were trained on a single 80GB A100 GPU for approximately 3 days. It takes 1.2 hours for these systems to process the evaluation dataset.

\section{Track 2: Language diarization}

\subsection{Methods}

In the language diarization track, we used the more conventional diarization technique, consisting of speech detection, segmentation into short overlapping windows, extraction of segment embeddings, and clustering of the segments, with VBx \cite{landini2022bayesian} based refinement of the initial clustering hypothesis.

As the first step in processing target speech data, segments containing speech were found from the recordings, using the Silero VAD model \cite{silero-vad}. Speech segments were further subsegmented, using a 5-second window with a 1-second shift. The use of 5-second window was inspired by the results from DISPLACE 2023 \cite{vachhani2023multi} and verified by our own initial experiments.

The resulting 5-second segments were processed by the language embedding model, which produces a 512-dimensional vector for each short segment. The backbone of the embeddings extractor is the Wav2Vec2-BERT model\footnote{\url{https://huggingface.co/facebook/w2v-bert-2.0}} shared by the Seamless4MT project \cite{seamless2023}. This model was pre-trained on 4.5M hours of unlabeled audio data covering more than 143 languages, using self-supervised loss. Wav2Vec2-BERT follows the same architecture as Wav2Vec2.0 \cite{baevski2020wav2vec}, but replaces the attention-block with a Conformer-block as introduced in \cite{gulati2020conformer}. It also uses mel-spectrogram representation of the audio as input, instead of the raw waveform. This particular Wav2Vec2-BERT model comprises 24 Conformer layers with approximately 600M parameters. The Wav2Vec2-BERT  model was converted into a language identification model by feeding its outputs through an attentive pooling layer, a fully connected layer with ReLU and BatchNorm, and the final output layer, corresponding to the languages of the training set. The model is trained using cross-entropy loss on random 2 to 4-second chunks of language-labeled training data. Point source noises and simulated room impulse responses (RIRs) from the SLR28 Room Impulse Response and Noise Database \cite{ko2017study} were used for on-the-fly data augmentation. Segment embeddings are extracted from the output of the first dense layer after the pooling layer. Low-rank adaptation (LoRA) \cite{hu2021lora} is used for finetuning the pre-trained Wav2Vec2-BERT model, with $\texttt{rank}=32$, $\alpha=32$ and $\texttt{dropout}=0.05$. Supervised training was performed using an effective batch size of 64, peak learning rate $10^{-3}$ and weight decay $10^{-3}$.  Due to the use of LoRA, the number of trainable parameters in the model is only 7.9M.

\begin{table}[tb]
\centering
\caption{Amount of training data per language for training the language embedding model for Track 2.}
\label{table:languages_numbers}
\setlength{\tabcolsep}{3pt}   
\begin{tabular}{lr|lr|lr}
\toprule
Language & Hours & Language & Hours & Language & Hours \\ \midrule
Amharic & 5.4 & Haiti Creole & 4.4 & Russian & 30.4 \\
Arabic & 13.4 & Hausa & 5.3 & Spanish & 26.0 \\
Azerbaijani & 5.0 & Hindi & 32.9 & Swahili & 5.4 \\
Belarusian & 4.9 & Indonesian & 1.7 & Tagalog & 6.0 \\
Bengali & 9.8 & Italian & 4.6 & Tamil & 8.6 \\
Bosnian & 4.8 & Japanese & 27.1 & Thai & 35.2 \\
Bulgarian & 5.1 & Khmer & 0.1 & Tibetan & 5.0 \\
Cantonese & 7.6 & Korean & 27.6 & Tigrinya & 0.0 \\
Chinese & 113.9 & Lao & 0.1 & Turkish & 5.4 \\
Croatian & 5.1 & Pashto & 5.4 & Ukrainian & 5.3 \\
English & 646.3 & Persian & 18.0 & Urdu & 11.6 \\
French & 9.3 & Portuguese & 5.4 & Uzbek & 5.8 \\
Georgian & 5.5 & Punjabi & 0.7 & Vietnamese & 29.5 \\
German & 5.6 & Romanian & 5.4 &  &  \\ \bottomrule
\end{tabular}
\end{table}

We tried various datasets for training the language embedding model. Initial experiments with the VoxLingua107 dataset \cite{valk2021slt} gave poor results on DISPLACE data (see section \ref{track2_analysis}).  Therefore, we opted to use data from NIST Language Recognition Evaluations (LREs) and Speaker Recognition Evaluations (SREs) for training the embedding model. Specifically, the language embeddings extractor was trained on NIST LRE 2003 evaluation data (LDC2006S31),  NIST LRE 2005 evaluation data (LDC2008S05),  NIST LRE 2007 evaluation data (LDC2009S04), NIST LRE 2009 evaluation data (LDC2014S06), NIST LRE 2007 training data (LDC2009S05), NIST SRE 2008 training data (LDC2011S05). Those datasets contain mostly conversational telephone speech, including English with Indian accent.   
The amount of speech data per language is given in Table \ref{table:languages_numbers}. 

Although the languages used in  DISPLACE~2024 Track 2 development and evaluation data were not known during the challenge period, the DISPLACE 2023 \cite{baghel23_interspeech} report suggests that they could include Indian-accented English, Hindi, Telugu, Bangla/Bengali, Kannada, Tamil. Table \ref{table:languages_numbers} shows that Telugu and Kannada were not covered by the training data used for training language embeddings. In order to adapt the embeddings to the DISPLACE 2024 scenario, we fine-tuned the embeddings model on development data from Track 3 which has been segmented and transcribed according to the language. This gives us around 3.5 hours of in-domain data (see Table \ref{tab:track3-durs}). Fine-tuning was performed for 6 epochs from the checkpoint trained on 10 epochs of NIST data, using a learning rate schedule where the peak learning rate is 10 times smaller than when training the initial model.

The 5-second segments were clustered using a language recognition model based on a LDA/PLDA, trained on Track 3 development data. The LDA/PLDA model transforms centered language embeddings to 150 dimensions using LDA and estimates a PLDA model on the length-normalized features. The LDA/PLDA model is used to evaluate the cross-similarity across all 5-second segment pairs in each target recording. The similarities are used to perform initial clustering of the 5-second segments, using AHC. The initial language segmentation is finally refined using Bayesian HMM clustering (VBx) \cite{landini2022bayesian}, using the following parameters: $P_{\texttt{loop}}=0.9$, $F_{a}=9$, $F_b=4$.

\begin{figure*}[t]
    \centering
    \includegraphics[width=1\textwidth]{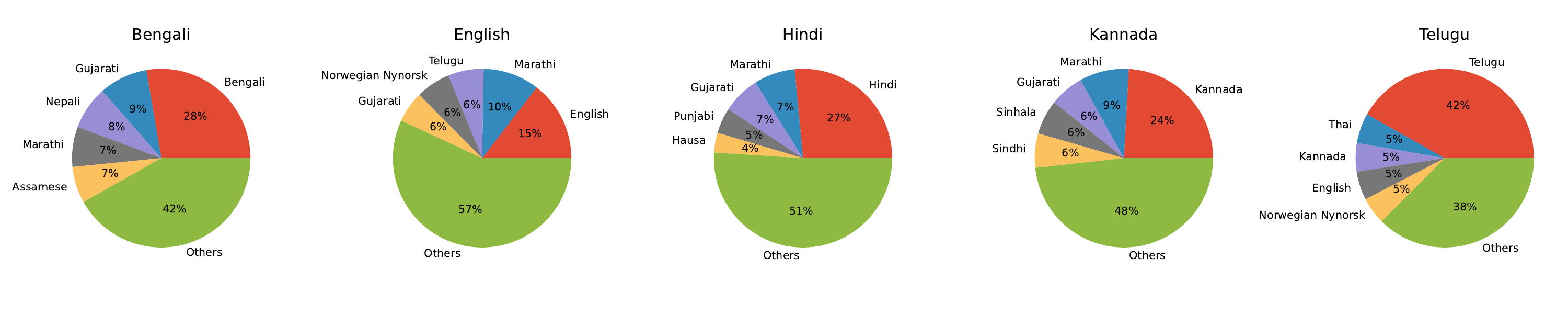}
    \caption{Top 5 most frequent predicted languages for Track 3 development utterances in the five given languages, based on the language identification model trained on VoxLingua107.}
    \label{fig:vl107-track3}
\end{figure*}

\begin{table}[tb]
\caption{Amount of data per language in Track 3 development data.}
\label{tab:track3-durs}
\centering
\begin{tabular}{lr}
\toprule
Language & Amount (hh:mm) \\
\midrule
Bengali  & 0:26   \\
Hindi    & 0:24   \\
English  & 1:47   \\
Kannada  & 0:12   \\
Telugu   & 0:37  \\
\bottomrule
\end{tabular}
\end{table}

\subsection{Results}

\begin{table}[tb]
\caption{Language diarization error rates (DER) on Track 2 development and evaluation data, using different model configurations.}
\label{tab:track2-results}
\setlength{\tabcolsep}{2pt}   
\begin{tabular}{p{2.6cm}|c|c|c|r}
\toprule
Training data for\\embeddings & LDA/PDA & VBx & Dev (conf. int.)  & Eval \\
\midrule
\multicolumn{5}{l}{\textit{\textbf{Baselines}}} \\ \midrule
\multicolumn{3}{l|}{No segmentation}           & 44.4 (41.9-49.1) & \\
\multicolumn{3}{l|}{DISPLACE 2023 baseline} & 48.6 (46.3-52.8) & \\
\multicolumn{3}{l|}{DISPLACE 2024 baseline} & 40.7 (37.9-45.3) &  32.7 \\ \midrule
VL107                        & VL107 & \xmark & 38.3 (35.7-43.4)  & \\                                           
VL107                        & Track3 dev & \xmark & 32.9 (30.3-37.5) & \\
VL107  + Track3 dev         & Track3 dev & \xmark & 30.1 (27.7-34.8) & \\
NIST                                & NIST      & \xmark & 30.9 (28.4-35.6)  &      \\
NIST                                & NIST      & \cmark & 29.7 (27.6-34.9)  &      \\
NIST                                & Track3 dev     & \xmark  & 31.3 (28.9-36.3) &      \\
NIST                               & Track3 dev    & \cmark & 28.7 (26.1-33.5) & 29.6 \\
NIST + Track3 dev                             & Track3 dev      & \xmark  & 29.3 (26.8-34.2) &      \\
NIST + Track3 dev                             & Track3 dev      & \cmark & 28.2 (25.6-33.0) & \textbf{27.6} \\
\bottomrule
\end{tabular}
\end{table}

Table \ref{tab:track2-results} presents the performance of various baseline systems and our own models on the development and evaluation datasets for Track 2. Confidence intervals \cite{Confidence_Intervals} on development data are computed by treating each recording as IID.
Notably, the DISPLACE 2023 baseline, which uses an EPACA-TDNN model trained on VoxLingua107 dataset for generating language embeddings, followed by the clustering of short segments using AHC, does not outperform the simplistic baseline that attributes all speech to a single language. However, the DISPLACE 2024 baseline that substitutes the EPACA-TDNN language embeddings with language detection posterior probabilities derived from Whisper, and incorporates VBx into the clustering step, achieves an improvement over the ``uninformative'' baseline.

The results further indicate that language embeddings trained using data from NIST LREs and SREs significantly outperform those trained with VoxLingua107 (VL107) data for the DISPLACE 2024 dataset. However, substantial gains are observed when in-domain data from Track 3 is utilized for estimating the LDA/PLDA model and for finetuning the embeddings. This approach not only enhances the performance of the VoxLingua107 based model but also narrows the gap to the models trained on NIST datasets. The system corresponding to the last line in the table obtained the best results on evaluation data among all teams.

\subsection{Analysis}
\label{track2_analysis}

Our investigation revealed that the VoxLingua107 dataset, effective for various language recognition tasks, showed weak performance on the DISPLACE 2024 dataset.  To decipher the underlying causes of this problem,  we assessed the language identification capabilities of a model trained on VoxLingua107 using the Track 3 development dataset, evaluating it through its posterior probabilities without employing LDA/PLDA postprocessing. Although the model achieved an accuracy of 95.4\% on the VoxLingua107 development dataset, its performance dramatically decreased to 22.2\% on the Track 3 dataset. Our analysis, shown in Figure \ref{fig:vl107-track3}, identified a trend across languages: while the correct language was often identified, recall rates were significantly low, from 15\% for English to 42\% for Telugu. This drop in accuracy can be attributed to factors like environmental noise and the conversational speech style. However, a major reason for the decline was the inclusion of non-native speech in the DISPLACE 2024 data. Prior study has shown that models trained on VoxLingua107 face dramatic accuracy losses with non-native accents \cite{kukk2022improving}, and that such models could be improved by also using a  lexicon-free character-based speech recognition for various languages to transcribe speech, followed by applying a text-based classification model on these transcripts. The combined model approach could potentially enhance language diarization and segmentation tasks as well.

\subsection{Runtime performance}

Training of the language embedding model was performed on 6 P100 GPUs and it took approximately 4 hours. Finetuning the model on Track 3 data takes a few minutes on one GPU.
Processing test data from start to finish takes about 0.08 $\times$ realtime, assuming one GPU and one CPU.

\section{Conclusion}

This work presents our submissions to the DISPLACE 2024 challenge. For the speaker diarization track, our best system combines \textit{pyannote.audio} speaker diarization pipelines where the segmentation is done either by a model trained with a powerset objective function or by a joint separation-diarization model trained with our recently proposed PixIT loss. The latter system is improved upon by extracting speaker embeddings directly from local separated sources. The ensemble reaches a DER of 27.1\% on the phase one evaluation data. In the language diarization track, a 27.6\% DER score is achieved by combining local language embeddings from a pre-trained Wav2Vec2-BERT model with clustering using AHC and VBx, based on similarity scores from LDA/PLDA. Our systems achieved first places in both of the tracks.

\section{Acknowledgements}

The research reported in this paper was supported by the Agence de l'Innovation Défense under the grant number 2022~65~0079, and by the Estonian Centre of Excellence in AI.
This work was granted access to the HPC resources of GENCI-IDRIS under the allocations AD011014274.

\bibliographystyle{IEEEtran}
\bibliography{mybib}

\end{document}